\documentclass[showpacs,preprintnumbers,nofootinbib]{revtex4}
\usepackage{amssymb}
\usepackage{amsmath}
\usepackage{graphicx}
\usepackage{dcolumn}
\usepackage{bm}

\begin{document}

\title{The quantization of the $B=1$ and $B=2$ Skyrmions}
\author{Jimmy Fortier and Luc Marleau}
\email{lmarleau@phy.ulaval.ca}
\affiliation{D\'epartement de Physique, de G\'enie Physique et d'Optique, Universit\'e
Laval, Qu\'ebec, Qu\'ebec, Canada G1K 7P4}
\date{\today}

\begin{abstract}
We propose to set the Skyrme parameters $F_{\pi }$ and $e$ such that they
reproduce the physical masses of the nucleon and the deuteron. We allow
deformation using an axially symmetric solution and simulated annealing to
minimize the total energy for the $B=1$ nucleon and $B=2$ deuteron. First  we find that axial deformations
are responsible for a significant  reduction (factor of \(\approx4\)) of     the rotational energy but
also that it is not possible to get a common set of parameters $F_{\pi }$ and $e$
which would fit both nucleon and deuteron masses simultaneously at least for
$m_{\pi }=$ 138 MeV, 345 MeV and 500 MeV. This suggests that either $%
m_{\pi}>500$ MeV or additional terms must be added to the Skyrme Lagrangian.
\end{abstract}

\pacs{12.39.Dc, 10.11Lm}
\maketitle

\section{\label{sec:Intro}Introduction}

The Skyrme model \cite{Skyrme:1961vq} is a nonlinear theory of pions that
admits topological solitons solutions called Skyrmions. These solutions fall
into sectors characterized by an integer-valued topological invariant $B$.
In its quantized version, a Skyrmion of topological charge $B$ may be
identified with a nucleus with baryon number $B$.

Since the model is non-renormalizable, a canonical quantization is not
possible and one has to resort to semiclassical quantization of the zero
modes of the Skyrmion. This method adds kinematical terms (rotational,
vibrational,..) to the total energy of the Skyrmion. The $B=1$ Skyrmion was
first quantized by Adkins, Nappi and Witten \cite%
{Adkins:1983hy,Adkins:1983ya}. It provided then a useful mean to set the
parameters of the Skyrme model $F_{\pi }$ and $e$ by fitting to the proton
and delta masses. The experimental value of the pion mass $m_{\pi }$
completes the set of input parameters when a pion mass term is added to the
Skyrme Lagrangian. The same calibration was assumed by Braaten and Carson
\cite{Braaten:1988cc} in their quantization of the $B=2$ Skyrmions and their
predictions of a rather tightly bound and small sized deuteron. Further
analysis by Leese, Manton and Schroers \cite{Leese:1994hb} considering the
separation of two single Skyrmions in the most attractive channel led to
more accurate predictions for the deuteron.

Yet all these calculations have three caveats: First, they used a rigid-body
quantization, \emph{i.e.} the kinematic term is calculated from the solution
which minimizes the static energy neglecting any deformation that could
originate form the kinematical terms. This was pointed out by several
authors \cite{Braaten:1984qe,Rajaraman:1985ty} who proposed to improve the
solutions by allowing the $B=1$ Skyrmion to deform within the spherically
symmetric ansatz. Second, even with such deformations it was noted that the
set of parameters mentioned above does not allow for a stable spinning
solution for the delta since the rotational frequency $\Omega $ would not
satisfy the constraint for stability against pion emission, $\Omega ^{2}\leq
\frac{3}{2}m_{\pi }^{2}$. These two problems were addressed recently in \cite%
{Battye:2005nx,Houghton:2005iu}. Assuming an axial symmetry, the
calculations were performed using a simulated annealing algorithm allowing
for the minimization of the total energy (static and kinetic). It was also
shown that the stable spinning nucleon and delta masses could be obtained
only if the pion mass is fixed at more than twice its experimental value.
One may argue that in this case $m_{\pi }$ could be interpreted as a
renormalized pion mass which could explain its departure form the
experimental value. A third difficulty remains: Fixing the parameters of the
Skyrme model still involve the delta which is interpreted as a stable
spinning Skyrmion although physically it is an unstable resonance. Recently,
Manton and Wood \cite{Manton:2006tq} took a different approach and chose
data from $B=6$ Lithium-6 to set the Skyrme parameters. The purpose was to
provide for a better description of higher $B$ solutions. Unfortunately,
their calculations were based on a two-fold approximation, the rational map
ansatz and rigid-body quantization for values of $B$ up to 8.

In this work, we propose to set the Skyrme parameters such that they
reproduce the physical masses of the nucleon and the deuteron. We allow
deformation within an axially symmetric solution approximation and use
simulated annealing to minimize the total energy for the $B=1$ nucleon and $%
B=2$ deuteron. We argue that this procedure provide a solution which is very
close to if not the exact solution. Following the procedure in \cite%
{Battye:2005nx}, we find the sets of parameters $F_{\pi }$ and $e$ that are
required to fit the nucleon and deuteron masses respectively for $m_{\pi }=$
138 MeV, 345 MeV and 500 MeV. The numerical calculations are compared to
those obtained with the rational maps ansatz and rigid-body quantization. We
find the latter approximation to be misleading since it suggests that it is
possible to simultaneously fit the nucleon and deuteron masses which is not
the case when we perform an our numerical calculation even for larger $%
m_{\pi }=$ 500 MeV. However for the solution to remains realistic with
regard with the size of the nucleon or deuteron, lower values of $F_{\pi }$
and $e$ are to be excluded.

In section II, we introduce briefly the Skyrme model and find the static
energy for the axially symmetric solution proposed in \cite{Battye:2005nx}.
The quantization of rotational and isorotational excitation using this
solution leads to an expression for the kinetic energy terms at the end of
section III. These expression suggest that the axial symmetry could be
preserved to a large extent. Finally we discuss and compare our numerical
results from simulated annealing with an axial solution on a two dimensional
grid with that coming from rational maps with rigid-body approximation in
the last section.

\section{\label{sec:Skyrme}The Skyrme model}

The $SU(2)$ Skyrme Lagrangian density is
\begin{equation}
\mathcal{L}_{S}=-\frac{{F_{\pi }^{2}}}{{16}}\mathrm{Tr}\left( {L_{\mu
}L^{\mu }}\right) +\frac{1}{{32e^{2}}}\mathrm{Tr}\left[ {L_{\mu },L_{\nu }}%
\right] ^{2}+\frac{{m_{\pi }^{2}F_{\pi }^{2}}}{{8}}\left( {\mathrm{Tr}U-2}%
\right)  \label{eq:Skyrme_lagr}
\end{equation}%
where $L_{\mu }$ is the left-handed chiral current $L_{\mu }=U^{\dag
}\partial _{\mu }U$ and the parameters $F_{\pi }$ and $e$ are respectively
the pion decay constant and the dimensionless Skyrme constant. $U$ is a $%
SU(2)$ field associated to the pion field $\pi $ by
\begin{equation}
U=\sigma +i\boldsymbol{\tau }\cdot \boldsymbol{\pi }  \label{eq:U}
\end{equation}%
where $\boldsymbol{\tau }$ are the Pauli matrices, $\boldsymbol{\pi }=(\pi
_{1},\pi _{2},\pi _{3})$ is the triplet of pion fields and the scalar meson
field $\sigma $ satisfy $\sigma ^{2}+\boldsymbol{\pi }\cdot \boldsymbol{\pi }%
=1$. The third term where $m_{\pi }$ is the pion mass was first added by
Adkins and Nappi \cite{Adkins:1983hy} to account for the chiral symmetry
breaking observed in nature.

The field configurations that satisfy the boundary condition
\begin{equation}
U(\mathbf{r},t)\rightarrow \mathbf{1}\hspace{1cm}\text{for}\hspace{0.3cm}|%
\mathbf{r}|\rightarrow \infty  \label{eq:finitude}
\end{equation}%
fall into topological sectors labelled by a topological invariant
\begin{equation}
B=\frac{1}{{2\pi ^{2}}}\int {d^{3}x\det \left\{ {L_{i}^{a}}\right\} }=-\frac{%
{\varepsilon ^{ijk}}}{{48\pi ^{2}}}\int {\mathrm{d}^{\mathrm{3}}x}\mathrm{Tr}%
\left( {L_{i}[L_{j},L_{k}]}\right)  \label{eq:charge}
\end{equation}%
taking integral values. Skyrme interpreted this topological invariant as the
baryon number. The minimal static energy Skyrmion for $B=1$ and $B=2$ turns out to
have spherical and axial symmetry respectively. Since we are only interested
by these values of $B$, the solution will be cast in terms of
cylindrical coordinates $(\rho ,\theta ,z)$ in the form
\begin{equation}
\sigma =\psi _{3}\hspace{1cm}\pi _{1}=\psi _{1}\cos n\theta \hspace{1cm}\pi
_{2}=\psi _{1}\sin n\theta \hspace{1cm}\pi _{3}=\psi _{2}
\label{eq:solaxiale}
\end{equation}%
introduced in \cite{Krusch:2004uf} where $\boldsymbol{\psi }(\rho ,z)=(\psi
_{1},\psi _{2},\psi _{3})$ is a three-component unit vector that is
independent of the azimuthal angle $\theta $. The boundary conditions (\ref%
{eq:finitude}) implies that $\boldsymbol{\psi }\rightarrow \left(
0,0,1\right) $ as $\rho ^{2}+z^{2}\rightarrow \infty $. Moreover, we must
impose that $\psi _{1}=0$ and $\partial _{\rho }\psi _{2}=\partial _{\rho
}\psi _{3}=0$ at $\rho =0$. The $B=1$ hedgehog solution appears as a special
case of (\ref{eq:solaxiale}) having spherical symmetry and corresponds to
\begin{equation*}
(\psi _{1},\psi _{2},\psi _{3})=(\sin F\sin \theta ,\sin F\cos \theta ,\cos
F)
\end{equation*}%
where $F=F(r)$ is the profile or chiral angle.

With the axial ansatz (\ref{eq:solaxiale}) and a appropriate scaling
\footnote{%
We have used ${2\sqrt{2}}/{eF_{\pi}}$ and ${F_{\pi}}/{2\sqrt{2}}$ as units
of length and energy respectively.} the expressions for the static energy
and the baryon number become
\begin{equation}  \label{eq:Estataxiale}
\begin{split}
E_n&=-\int\text{d}^3x\mathcal{L}_S \\
&=2\pi\left(\frac{F_\pi}{2\sqrt{2}e}\right)\int\text{d}z\text{d}\rho\rho%
\biggl\{\left(\partial_\rho\boldsymbol{\psi}\cdot\partial_\rho\boldsymbol{%
\psi} +\partial_z\boldsymbol{\psi}\cdot\partial_z\boldsymbol{\psi}%
\right)\left(1+n^2\frac{\psi^2_1}{2\rho^2}\right) +\frac{1}{2}|\partial_z
\boldsymbol{\psi}\times\partial_\rho \boldsymbol{\psi}|^2+n^2\frac{\psi^2_1}{%
\rho^2}+2\beta^2\left(1-\psi_3\right)\biggr\}
\end{split}%
\end{equation}

\begin{equation}
B=\frac{n}{\pi }\int \text{d}z\text{d}\rho \,\psi _{1}|\partial _{\rho }%
\boldsymbol{\psi }\times \partial _{z}\boldsymbol{\psi }|  \label{eq:Bcyl}
\end{equation}%
with $\beta =\frac{2\sqrt{2}m_{\pi }}{eF_{\pi }}$.

However, to describe baryons, Skyrmions must acquire a well defined spin and
isospin state. This is possible only upon proper quantization of the
Skyrmions as it will be done in the next section. As we shall see in the
next section adding (iso-)rotational energy will in general brake the axial
symmetry manifest in $B=1$ and $B=2$ static solutions.

\section{\label{sec:Quantization}Quantization}

Since the Skyrme Lagrangian (\ref{eq:Skyrme_lagr}) is invariant under
rotation and isorotation \footnote{%
Since we are interested only in the computation of the static properties, we
will ignore translational modes and quantize the Skyrmions in their rest
frame.}, the usual method of Skyrmion quantization consist of allowing the
zero modes to depend on time and then quantize the resulting dynamical
system according to standard semiclassical methods. From this perspective,
the dynamical ansatz is assumed to be
\begin{equation}
\widetilde{U}(\mathbf{r},t)=A_{1}(t)U(R(A_{2}(t)\mathbf{r})A_{1}^{\dag }(t)
\label{eq:Urot}
\end{equation}%
where $A_{1}$, $A_{2}$ are $SU(2)$ matrices and $R_{ij}(A_{2})=\frac{1}{2}%
\text{Tr}(\tau _{i}\,A_{2}\tau _{j}\,A_{2}^{\dag })$ is the associated $%
SO(3) $ rotation matrix. Introducing this ansatz into the Skyrme Lagrangian (%
\ref{eq:Skyrme_lagr}) one gets the kinematical contribution to the total
energy which can be cast in the form
\begin{equation}
T=\frac{1}{2}\,a_{i}\,U_{ij}\,a_{j}-a_{i}\,W_{ij}\,b_{j}+\frac{1}{2}%
\,b_{i}\,V_{ij}\,b_{j}  \label{eq:Trot}
\end{equation}%
where $a_{j}=-i\text{Tr}\left( \tau _{j}\,A_{1}^{\dag }\dot{A_{1}}\right) $,
$b_{j}=i\text{Tr}\left( \tau _{j}\,\dot{A_{2}}A_{2}^{\dag }\right) $ and the
$U_{ij},V_{ij}$ and $W_{ij}$ are inertia tensors
\begin{equation}
U_{ij}=-\left( \frac{2\sqrt{2}}{e^{3}F_{\pi }}\right) \int \text{d}^{3}x%
\biggl\{\text{Tr}\left( T_{i}T_{j}\right) +\frac{1}{8}\text{Tr}\left( \left[
L_{k},T_{i}\right] \left[ L_{k},T_{j}\right] \right) \biggr\},
\label{eq:Uij}
\end{equation}%
\begin{equation}
V_{ij}=-\left( \frac{2\sqrt{2}}{e^{3}F_{\pi }}\right) \epsilon
_{ikl}\epsilon _{jmn}\int \text{d}^{3}xx_{k}x_{m}\biggl\{\text{Tr}\left(
L_{l}L_{n}\right) +\frac{1}{8}\text{Tr}\left( \left[ L_{p},L_{l}\right] %
\left[ L_{p},L_{n}\right] \right) \biggr\},  \label{eq:Vij}
\end{equation}%
\begin{equation}
W_{ij}=\left( \frac{2\sqrt{2}}{e^{3}F_{\pi }}\right) \epsilon _{jkl}\int
\text{d}^{3}xx_{k}\biggl\{\text{Tr}\left( T_{i}L_{l}\right) +\frac{1}{8}%
\text{Tr}\left( \left[ L_{m},T_{i}\right] \left[ L_{m},L_{n}\right] \right) %
\biggr\}  \label{eq:Wij}
\end{equation}%
where $T_{i}=iU^{\dag }\left[ \frac{\tau _{i}}{2},U\right] $. For the axial
ansatz (\ref{eq:solaxiale}), these tensors are all diagonal and satisfy $%
U_{11}=U_{22}$, $V_{11}=V_{22}$, $W_{11}=W_{22}$ and $U_{33}=\frac{W_{33}}{n}%
=\frac{V_{33}}{n^{2}}$. The components of these inertia tensors are
\begin{equation}
\begin{split}
U_{11}=2\pi \left( \frac{2\sqrt{2}}{e^{3}F_{\pi }}\right) \int \text{d}z%
\text{d}\rho \rho \biggl\{\psi _{1}^{2}+2\psi _{2}^{2}+\frac{1}{2}\biggl[&
\left( \partial _{\rho }\boldsymbol{\psi }\cdot \partial _{\rho }\boldsymbol{%
\psi }+\partial _{z}\boldsymbol{\psi }\cdot \partial _{z}\boldsymbol{\psi }%
+n^{2}\frac{\psi _{1}^{2}}{\rho ^{2}}\right) \psi _{2}^{2}+\left( \partial
_{\rho }\psi _{3}\right) ^{2}+\left( \partial _{z}\psi _{3}\right) ^{2} \\
& +n^{2}\frac{\psi _{1}^{4}}{\rho ^{2}}\biggr]\biggr\},
\end{split}
\label{eq:U11cyl}
\end{equation}

\begin{equation}  \label{eq:U33cyl}
U_{33}=2\pi\left(\frac{2\sqrt{2}}{e^3F_\pi}\right)\int \text{d}z\text{d}%
\rho\rho\psi^2_1\left(\partial_\rho\boldsymbol{\psi}\cdot\partial_\rho%
\boldsymbol{\psi} +\partial_z\boldsymbol{\psi}\cdot\partial_z\boldsymbol{\psi%
}+2\right),
\end{equation}

\begin{equation}  \label{eq:V11cyl}
V_{11}=2\pi\left(\frac{2\sqrt{2}}{e^3F_\pi}\right)\int \text{d}z\text{d}%
\rho\rho\biggl\{|\rho\partial_z\boldsymbol{\psi}-z\partial_\rho\boldsymbol{%
\psi}|^2\left(1 +n^2\frac{\psi^2_1}{2\rho^2}\right) +z^2n^2\frac{\psi^2_1}{%
\rho^2}+\frac{1}{2}\left(\rho^2+z^2\right) |\partial_\rho\boldsymbol{\psi}%
\times\partial_z\boldsymbol{\psi}|^2\biggr\},
\end{equation}

\begin{equation}
\begin{split}
W_{11}& =2\pi \left( \frac{2\sqrt{2}}{e^{3}F_{\pi }}\right) \int \text{d}z%
\text{d}\rho \rho \biggl\{\left[ \psi _{1}\left( \rho \partial _{z}\psi
_{2}-z\partial _{\rho }\psi _{2}\right) -\psi _{2}\left( \rho \partial
_{z}\psi _{1}-z\partial _{\rho }\psi _{1}\right) \right] \biggl(1+\frac{1}{2}%
\left[ \left( \partial _{z}\psi _{3}\right) ^{2}+\left( \partial _{\rho
}\psi _{3}\right) ^{2}+\frac{\psi _{1}^{2}}{\rho ^{2}}\right] \biggr) \\
& \phantom{=2\pi\left(\frac{2\sqrt{2}}{e^3F_\pi}\right)\int
\text{d}z\text{d}\rho\rho\biggl\{}+\frac{\psi _{3}}{2}\left( z\partial
_{z}\psi _{3}+\rho \partial _{\rho }\psi _{3}\right) \left[ \partial _{\rho
}\psi _{2}\partial _{z}\psi _{1}-\partial _{\rho }\psi _{1}\partial _{z}\psi
_{2}\right] \\
& \phantom{=2\pi\left(\frac{2\sqrt{2}}{e^3F_\pi}\right)\int
\text{d}z\text{d}\rho\rho\biggl\{}+\frac{z\psi _{1}\psi _{2}}{2\rho }\left(
2+\partial _{\rho }\boldsymbol{\psi }\cdot \partial _{\rho }\boldsymbol{\psi
}+\partial _{z}\boldsymbol{\psi }\cdot \partial _{z}\boldsymbol{\psi }%
\right) \biggr\}.
\end{split}
\label{eq:W11cyl}
\end{equation}%
Let us note that $W_{11}\neq 0$ only for $n=1$. In order to obtain the
energy corresponding to the nucleons and the deuteron, we must compute the
Hamiltonian for the rotational and isorotational degrees of freedom in term
of the inertia tensors (\ref{eq:U11cyl}) to (\ref{eq:W11cyl}) as well as its
eigenvalues for the states corresponding to these particles.

The body-fixed isospin and angular momentum canonically conjugate to $%
\mathbf{a}$ and $\mathbf{b}$ are respectively
\begin{equation}
K_{i}=\frac{\partial T}{\partial a_{i}}=U_{ij}a_{j}-W_{ij}b_{j},
\label{eq:Kiso}
\end{equation}%
\begin{equation}
L_{i}=\frac{\partial T}{\partial b_{i}}=-W_{ij}^{\text{T}}a_{j}+V_{ij}b_{j}.
\label{eq:Lrot}
\end{equation}%
These momenta are related to the usual space-fixed isospin ($\mathbf{I}$)
and spin ($\mathbf{J}$) by the orthogonal transformations
\begin{equation}
I_{i}=-R(A_{1})_{ij}K_{j},  \label{eq:I}
\end{equation}%
\begin{equation}
J_{i}=-R(A_{2})_{ij}^{\text{T}}L_{j}.  \label{eq:J}
\end{equation}

We can now write the Hamiltonian for the rotational and isorotational
degrees of freedom as
\begin{equation*}
H=\mathbf{K}\cdot \mathbf{a}+\mathbf{L}\cdot \mathbf{b}-T
\end{equation*}%
where $T$ is the kinematical energy (\ref{eq:Trot}).

The quantization procedure consist of promoting four-momenta as quantum
operators that satisfy each one the $SU(2)$ commutation relations. According
to (\ref{eq:I}) and (\ref{eq:J}), we see that the Casimir invariants satisfy
$\mathbf{I}^{2}=\mathbf{K}^{2}$ and $\mathbf{J}^{2}=\mathbf{L}^{2}$. Then
the operators form a $O(4)_{\text{I},\text{K}}\otimes O(4)_{\text{L},\text{J}%
}$ Lie algebra. The physical states on which these operators act are the
states formed in the base $|\Psi \rangle =|ii_{3}k_{3}\rangle
|jj_{3}l_{3}\rangle $ where $-i<i_{3}$, $-j<j_{3}$ and $l_{3}<j$ that
satisfy the constraints formulated by Finkelstein and Rubinstein \cite%
{Finkelstein:1968hy}. One of these constraints, namely,

\begin{equation}
e^{2\pi i\mathbf{n}\cdot \mathbf{L}}|\Psi \rangle =e^{2\pi i\mathbf{n}\cdot
\mathbf{K}}|\Psi \rangle =(-1)^{B}|\Psi \rangle ,  \label{eq:FR1}
\end{equation}
implies that the spin and isospin must be an integer for even $B$ or a
half-integer for odd baryon numbers. Another constraint
\begin{equation}
\left( nK_{3}+L_{3}\right) |\Psi \rangle =0  \label{eq:FR2}
\end{equation}%
comes from the axial symmetry imposed on the solution here.

This formalism allows one to obtain the total energy in terms of the inertia
tensors of (\ref{eq:U11cyl}-\ref{eq:W11cyl}) for the nucleon \cite%
{Houghton:2005iu}
\begin{eqnarray}
E_{N}  &=&E_{1}+\frac{1}{4}\left[ \frac{\left( 1-\frac{W_{11}}{U_{11}}\right) ^{2}}{%
V_{11}-\frac{W_{11}^{2}}{U_{11}}}+\frac{1}{U_{11}}+\frac{1}{2U_{33}}\right]
 \label{eq:Enucleons}\end{eqnarray}%
and for the deuteron \cite{Braaten:1988cc}
\begin{eqnarray}
E_{D} &=&E_{2}+\frac{1}{V_{11}}.\label{eq:Edeuteron}
\end{eqnarray}

For the $B=1$ solution the minimization of the static energy $E_{1}$ leads
to a spherically symmetric solution. The nucleon mass in (\ref{eq:Enucleons}%
) however receives (iso-)rotational energy contributions from
(iso-)rotations around the three principal axis among which two are
equivalent (direction 1 and 2). The deformation should then lead to a
axially symmetric solution along the $z$-axis (direction 3) as argued in Refs.\cite%
{Houghton:2005iu,Battye:2005nx}.

The situation is somewhat different for $B=2$ which is known to have a
toroidal solution upon minimization of the static energy $E_{2}$. In our
scheme it corresponds to the axially symmetric solution with respect to the $%
z$-axis. Assuming such a solution the deuteron mass in (\ref{eq:Edeuteron})
gets kinetic energy contributions which may be recast in the form
\begin{equation}
E_\text{rot} =\frac{1_{}^{}}{2V_{11}}+\frac{1}{2V_{22}}=\frac{1}{V_{11}},
\label{eq:Tdeuteron}
\end{equation}%
which means that the contributions come from rotations perpendicular to the
axis of symmetry. Unfortunately there is no guaranty that in a non-rigid
rotator approach the axial symmetry will be preserved when the rotational
terms are added. However the deformations are not expected to be very large
since the magnitude rotational energy only accounts for less than 4\% of the
total mass of the deuteron in the rigid-body approximation \cite%
{Braaten:1988cc}. Allowing deformations should increase the moments of
inertia and bring the relative contribution of rotational energy to an even
lower value. Clearly large changes in the configuration are prohibited by an
eventual increase in the static energy $E_{2}$ so, from that argument alone
we can infer that deformations from axial symmetry are bound to be contained
into a 4\% effect. Our results will show that non-axial deformations
represents less than 1\% of the deuteron mass if they contribute at all.
This justifies the use of the axial solution in (\ref{eq:solaxiale}).

In order to obtain the configurations $\boldsymbol{\psi }(\rho ,z)=(\psi
_{1},\psi _{2},\psi _{3})$ that minimize the energies defined in (\ref%
{eq:Enucleons}) and (\ref{eq:Edeuteron}), we use a two-dimensional version
of a simulated annealing algorithm used in \cite{Longpre:2005fr}. The
minimization is carried out on a grid in a plane $(\rho ,z)$ made up of 250 $%
\times $ 500 points with a spacing of 0.042 \footnote{%
These parameters were adopted in order to be similar with those used in \cite%
{Houghton:2005iu,Battye:2005nx}}. The algorithm starts with an initial
configuration $\boldsymbol{\psi }_{0}(\rho ,z)$ on the grid and evolves
towards the exact solution. Here $\boldsymbol{\psi }_{0}(\rho ,z)$ is
generated using the suitable rational map ansatz \cite{Houghton:1997kg} to
ensure that the initial solution have the appropriate baryon number, $B=1$
or $2,$ with a profile function of the form
\begin{equation}
F(r)=4\arctan \left( e^{-\alpha r}\right)  \label{eq:fct_ini}
\end{equation}%
as inspired from Ref.\cite{Sutcliffe:1992bg}. Here $\alpha $ is a parameter
chosen so that the entire baryon number density (such that $B=1$ or $2$
respectively for the nucleon or the deuteron) fits into the grid.

\section{Results and discussion}

There are several ways to fix the parameters of the Skyrme model $e$, $F_\pi$
and $m_\pi$ and the authors of Ref. \cite{Battye:2005nx} showed that one
must take some precautions in order that the spinning solutions for the
nucleon and the delta remain stable against pion emission. And indeed it was
found that, in order to achieve a fit for the energies of spinning Skyrmions
to the masses of the nucleon and delta, one must impose a value for the pion
mass that is larger than its experimental value. Differences between the
fitted and experimental values of $F_\pi$ and $m_\pi$ are not proscribed
since after all, the values that enters the Lagrangian (\ref{eq:Skyrme_lagr}%
) are the unrenormalized parameters which could differ from the physical
ones.

Even so it remains that this procedure still assumes that a
physically unstable particle, the delta resonance, is described as
a stable spinning Skyrmion and to be perfectly consistent one
should instead rely on stable particles. For these reasons we
chose to carry out calculations using data from two stable
particles, the nucleon and the deuteron. We proceed as follows:
Assuming a value for the pion mass and $e$, and an initial value
for the Skyrme parameter $F_{\pi }$ we compute the lowest energy
solution for the spinning Skyrmion corresponding to the nucleon
using simulated annealing which leads to a mass prediction. We
iterate the procedure adjusting values of $F_{\pi }$ until the
predicted mass fits that of the nucleon. The same procedure is
repeated for the $B=2$ deuteron as well as for several values of
$e$. A set of points requires about two weeks of computer
calculations on a regular PC. Performing an equivalent calculation
on a 3D grid with similar spacing for a non-axial solution for
example would be prohibitive. This explain in parts why we use the
axially symmetric ansatz.

\begin{figure}[th]
\includegraphics[width=.6\textwidth]{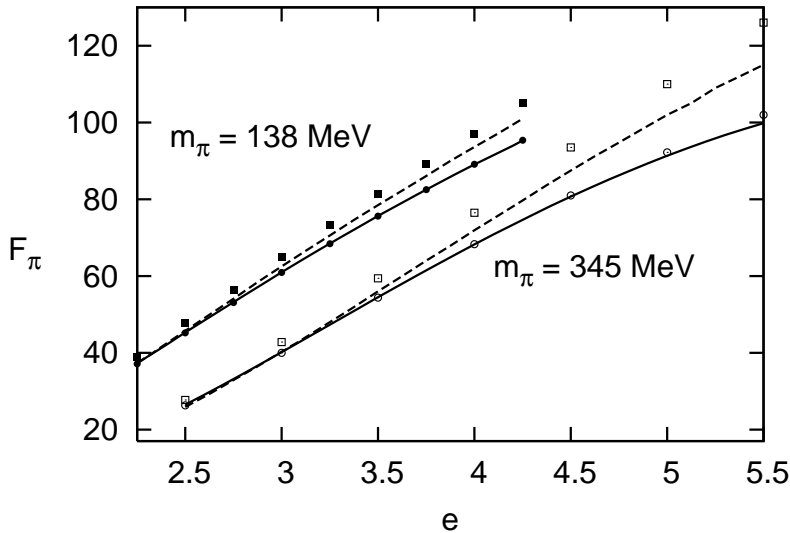}\newline
\caption{$F_{\protect\pi }$ as a function of $e$ for which $M+E_{\text{rot}}$
is equal to the nucleon mass (circles) and that of the deuteron (squares).
The solid and dashed lines correspond to the results obtained from the rigid
body approach for the nucleon and deuteron respectively. The set of data and
lines at the upper left are for $m_{\protect\pi }=138$ MeV whereas that at
the bottom right are for $m_{\protect\pi }=345$ MeV. }
\label{Fig:mpi}
\end{figure}

Initially, the first set of calculations was performed with the pion mass
equal to its experimental value $m_{\pi }=138$ MeV. Since Ref. \cite%
{Battye:2005nx} suggests that a pion mass value of $m_{\pi }=345$
MeV or more is necessary to avoid instability due to pion emission
we repeated the calculations and evaluated the Skyrme parameters
using $m_{\pi }=345$ MeV. The results are displayed in FIG.
\ref{Fig:mpi}. Although it may seem interesting here to consider
values of $e$ larger than those illustrated on FIG. \ref{Fig:mpi},
the physical relevance of lower values of $e$ is questionable.
Below a certain value, the rotational energy of the Skyrmion is
larger than the contribution coming from the pion mass term. As
was highlighted in Refs. \cite{Braaten:1984qe,Rajaraman:1985ty},
this leads to an unstable Skyrmion with respect to emission of
pions.

Our results for $B=1$ reproduce the same behavior as in Refs. \cite%
{Houghton:2005iu,Battye:2005nx}, \emph{i.e.} the deformation of
the Skyrmion becomes relatively important only for larger values
of $m_{\pi }$ and $e$. However, the results for $B=2$ are far more
interesting with respect to deformation. Indeed, as we can see
directly from FIG. \ref{Fig:mpi}, there is a noticeable difference
between our numerical results (squares) which allow for
deformations as long as they preserve axial symmetry and that of
the rigid body approximation (dashed lines). Note that our
implementation of the rigid body approximation shown here relies
on the rational map ansatz which is known to be accurate to a few
percent and so it could not be responsible alone such large
difference in energy. This difference indeed corresponds to a
surprisingly much smaller energy for the deformed spinning
Skyrmion than what is obtained from a the rigid body approximation
wether it is based on the rational map ansatz or not.

To
illustrate this difference, we carried out our
minimization procedure using the set of parameters $F_{\pi }=108$ MeV and $%
e=4.84$  for the deformed deuteron. The results, set SA, are listed in of Table \ref{tab:SAvsBraaten}.
For comparison we also present three  rigid body calculations: set SA-RB performed with our simulated annealing algorithm, set Ref.\cite{Braaten:1988cc}-RB  from  Braaten
et al and finally set RM-RB
obtained through the rational maps approximation.
Note that our results SA-RB agrees fairly well with that of Ref.\cite{Braaten:1988cc}.
\begin{table}[th]
\begin{tabular}{|lcccccc|}
\hline\hline
& \quad &\quad SA \quad&\quad SA-RB\quad&\quad Ref.\cite{Braaten:1988cc}-RB\quad
 & \quad  RM-RB \quad & \quad  Exp. \quad \\ \hline
$E_D$ (MeV) &  & 1679 & 1716  & 1720 & 1750 &  1876 \\
$E_\text{rot}$ (MeV) &  & 13.3  & 60.6 & 61.2 & 55.2 & - \\
$\langle r^2 \rangle_D^{1/2}$ (fm) &  & 0.94  & 0.93  & 0.92 & 0.94   & 2.095 \\ \hline
\end{tabular}%
\caption{Total energy, rotational energy and charge radius of the deuteron
using parameters  $F_{\pi }=108$ MeV and $%
e=4.84.$  The results are presented for our simulated annealing calculations with axial deformation (SA)
along with three rigid body calculations: with  simulated annealing (SA-RB),   from Braaten
et al. (Ref.\cite{Braaten:1988cc}-RB) and
with the rational maps approximation (RM-RB) respectively. The last column (Exp.) shows the experimental
values where  the charge radius come from \protect\cite{Ericson:1984}.}
\label{tab:SAvsBraaten}
\end{table}
Clearly  the mass of the deuteron $E_{D}$ must be larger than $E_{2}^{\text{%
min}}=1655$ MeV$,$ the minimum static energy for this choice of parameters. Computing the rotational energy for this solution leads to the rigid body approximation result (SA-RB)
for the deuteron mass of   $E_{D}=1716$ MeV$.$ As expected the results for the deformed deuteron (SA) $E_{D} =1679$ MeV
lies between these two values. Moreover our axial solution brings the
relative contribution of the rotational term to about 1\% of the minimum
static energy $E_{2}^{\text{min}}$. So   allowing for axial deformation reduces by about a factor of
4   the rotational term
going from 60.6 MeV to 13.3 MeV.  As for non-axial
deformations that might be present in a completely general solution, they must at the very most represent an
1\% correction to the deuteron mass since it is bounded by  $E_{2}^{\text{min}}$ $(=1655$ MeV) $<E_{D}^{\text{exact}%
}\leq $ $E_{D}$ $(=1679$ MeV). On the other hand they may still represent a significant
portion of the remaining 13.3 MeV rotational energy. We conclude nonetheless that the axial symmetry
ansatz represents a very good approximation of the deuteron configuration.
In addition, since these bounds are both based on axially symmetric
solutions and their energy only differ by 1\%, one can even contemplate the possibility
that the exact solution may have axial symmetry and therefore is to our solution, contrary to what physical
intuition might suggest. This has yet to be proven and needless to say that such a demonstration would require a 3D calculation with a level
of precision less than 1\%.

\begin{figure}[!ht]
\includegraphics[width=.6\textwidth]{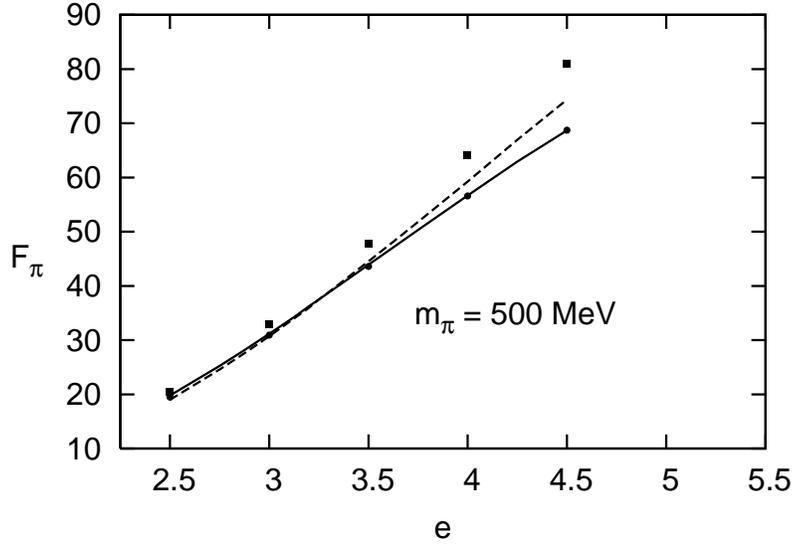}\newline
\caption{Same as FIG.\protect\ref{Fig:mpi} for $m_\protect\pi= 500$ MeV.}
\label{Fig:mpi500}
\end{figure}

Note that, both for $m_\pi=138$ MeV and 345 MeV, the solid and dashed lines
intersect for a relatively low value of $e. $ However despite what this
rigid body calculations suggests, there is no set $F_{\pi}$ and $e $ that
leads simultaneously to the masses of the nucleon and the deuteron. However,
the data of FIG. \ref{Fig:mpi} would indicate that the gap between the
values of $F_\pi$ decreases for a larger pion mass. For that reason, we
repeated the calculations with a relatively large $m_\pi=500$ MeV. Some of
the computations required that we adjust the spacing of our grid to 0.0057
for smaller $e$ since the configurations turned out to be much smaller in
size. The results for $m_\pi=500$ MeV are presented in FIG. \ref{Fig:mpi500}%
. The main conclusion one can draw is that even for this pion mass, not
common set of Skyrme parameters can be found.

\begin{figure}[!ht]
\includegraphics[width=.6\textwidth]{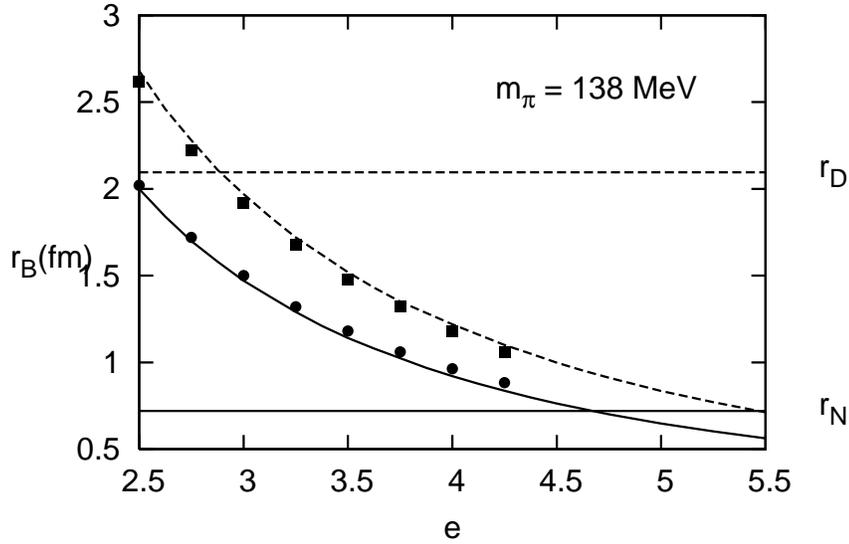}\newline
\caption{Charge radius of the nucleon (bold circles) and deuteron (bold
squares) for the set of parameters of FIG. \protect\ref{Fig:mpi} with $m_%
\protect\pi= 138$ MeV. The curves and the horizontal lines, solid for the
nucleons and dashed for the deuteron, correspond respectively to the results
obtained from the rigid body approach and to the experimental data. }
\label{Fig:rayon138}
\end{figure}

\begin{figure}[!ht]
\includegraphics[width=.6\textwidth]{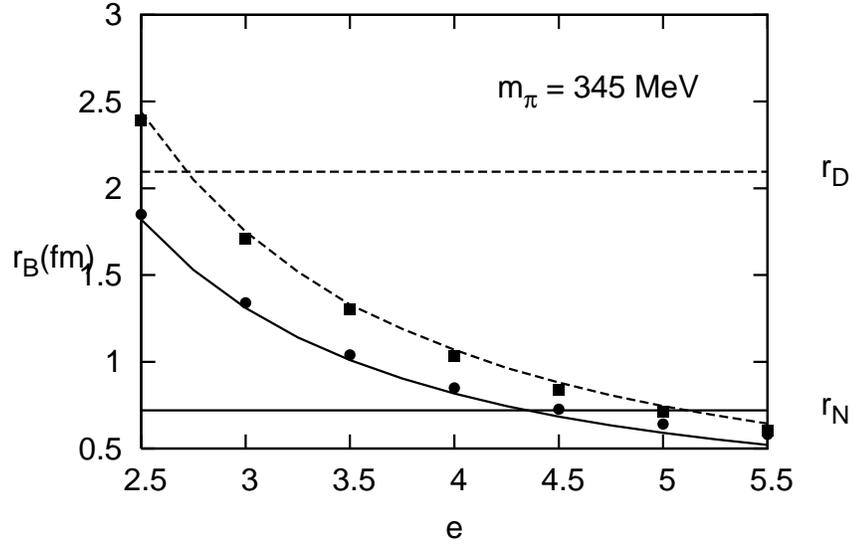}\newline
\caption{Same as FIG. \protect\ref{Fig:rayon138} for $m_\protect\pi= 345$
MeV.}
\label{Fig:rayon345}
\end{figure}

\begin{figure}[!ht]
\includegraphics[width=.6\textwidth]{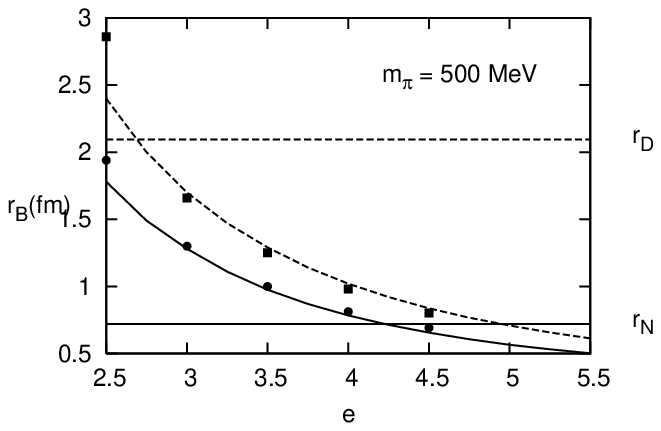}\newline
\caption{Same as FIG. \protect\ref{Fig:rayon138} for $m_\protect\pi= 500$
MeV.}
\label{Fig:rayon500}
\end{figure}

It is also interesting to note that the gap between the values of $F_{\pi }$
decrease as $e$ decreases. This would suggest that at very low values of $e$
a fit is possible. However computing the charge radius, \textit{i.e.} the
square root of
\begin{equation}
r_{n}^{2}=\frac{8}{e^{2}F_{\pi }^{2}\pi }\int
\text{d}z\text{d}\rho
\,\left( \rho ^{2}+z^{2}\right) \psi _{1}|\partial _{\rho }\boldsymbol{\psi }%
\times \partial _{z}\boldsymbol{\psi }|  \label{eq:chargeradius}
\end{equation}%
leads to an significant increase for the radius for small values of $e$ and $%
F_{\pi }$ as illustrated in FIGS. \ref{Fig:rayon138} to \ref{Fig:rayon500}.
So lower values of $e$ are incompatible with the physical size of the
deuteron and nucleon and may be discarded. The results also indicate that
the best fit for the radius of the nucleon and deuteron would favor
intermediate values of $e$ around $e\simeq 3.5$ while it looks fairly
insensitive to large changes in $m_{\pi }$.

To summarize, our calculations showed that the axial symmetry ansatz is a
very good approximation of the exact solution for the deuteron. This hints
at the possibility that it may even represent the exact solution. This
remains to be proved with a general 3D calculation. We also found that allowing for axial deformation reduces     the rotational energy by  a significant  factor.  On the other hand we
found that it is not possible to get a common set of parameters $F_{\pi }$
and $e$ which would fit both nucleon and deuteron masses simultaneously at
least for $m_{\pi }=$ 138 MeV, 345 MeV and 500 MeV. This conclusion should
hold even for the exact $B=2$ solution since if the solution was allowed to
adjust free of any symmetry contraints it would achieve a configuration with
lower total energy which would require larger values of $F_{\pi }$ for the
same set of $e.$ This suggests that either $m_{\pi }>500$ MeV or additional
terms must be added to the Skyrme Lagrangian. We also observed an increase
in the deformations due to the spinning of the $B=2$ Skyrmion (deuteron)
especially for larger values of $e$ and $F_{\pi }$ so the rigid body
approximation may not be appropriate in that case.

This work was were supported by the National Science and Engineering
Research Council.

\end{document}